\documentclass[proceedings]{JHEP3}

\PrHEP{PrHEP hep2001}                   
\conference{International Europhysics Conference on HEP}                

\usepackage{epsfig}                   

\def\beq{\begin{equation}}
\def\eeq{\end{equation}}

\def\beqn{\begin{eqnarray}}
\def\eeqn{\end{eqnarray}}
\def\sss{\scriptscriptstyle}
\def\mt{m_{\sss T}}
\def\pt{p_{\sss T}}
\def\ptq{p_{\sss T}^2}
\def\as{\alpha_{\sss S}}
\def\aem{\alpha_{em}}
\def\ep{\epsilon}
\def\abs#1{|#1|}

\title{Theory versus experiments in heavy flavour production}

\author{\speaker{Stefano Frixione}\\
        LAPP, Annecy, France and INFN, Sezione di Genova, Italy\\
        E-mail: \email{Stefano.Frixione@cern.ch}}

\abstract{I discuss the current status of the comparison between
theoretical predictions and experimental data, relevant to the 
production of open charm and bottom quarks in photon-hadron 
and photon-photon collisions. I advocate the use of a formalism
that matches fixed-order computations to resummed computations
in order to make firm statements on heavy flavour production 
as described by perturbative QCD.}

\begin{document}

\section{Introduction}

Heavy flavour physics has been traditionally a challenging testing
ground for the predictions of perturbative QCD. Loosely speaking,
we like to define a quark as heavy when its mass $m$ is much larger
than $\Lambda_{\sss QCD}$. This property entails the possibility of
computing in perturbation theory the cross section for the production
of an open heavy quark, which is not possible in the case of a light
quark. It is customary, although not always accurate, to say that the 
mass of the quark sets the hard scale of the production process, and
thus the relevant parameter to the perturbative expansion is $\as(m)$.
Furthermore, the condition $m\gg \Lambda_{\sss QCD}$ leads us to
expect that the perturbative predictions are only marginally
affected by power corrections and by contributions of non-perturbative
origin, that we can not compute from first principles.

There is no doubt that the top quark is a heavy quark; in fact, perturbative
QCD does a fairly good job in describing its production mechanism. There
is also a consensus on the fact that the bottom can be consistently
treated as a heavy flavour. The case of the charm is borderline; it is
difficult to list it together with u, d, and s quarks; on the other
hand, we expect non-perturbative physics to have a non-negligible
impact in this case, and we know that perturbative corrections are
huge, since $\as(1.5~{\rm GeV})\simeq 0.3$.

In this paper, I shall not deal with top physics, and I shall
concentrate on charm and bottom production in collider processes where
at least one of the incoming particles is an on-shell photon. Thus, I
shall not treat bottom production at hadronic colliders, nor charm and
bottom production at fixed-target experiments. I shall only briefly
remind the reader that Tevatron data for the production of $B$ mesons
lie above the QCD predictions obtained with ``default'' parameter
choices; however, QCD can predict the shape of $\pt$ spectrum,
and does well for a few $b\bar{b}$ correlations. Also, the comparison
between theory and data improves if $b$-jets are considered.  As far
as fixed-target charm hadroproduction is concerned, it appears that
perturbative QCD cannot reproduce the data, unless non-perturbative
effects, such as $\pt$-kick, are supplemented. Fortunately, these are 
moderate, and the agreement is, all in all, satisfactory. More details
can be found in ref.~\cite{Frixione:1997ma}

\section{Perturbative computations\label{sec:FOcomp}}

Using the factorisation theorem, I write the cross section for the
production of an inclusive open heavy quark $Q$ in photon-hadron
collisions as follows:
\beq
d\sigma_Q=\sum_j f^{(h)}_j \otimes d\hat{\sigma}_{\gamma j}
+\sum_{ij} f^{(\gamma)}_i \otimes f^{(h)}_j\otimes d\hat{\sigma}_{ij},
\label{factphoto}
\eeq
where $f^{(\gamma)}_i$ and $f^{(h)}_j$ are the parton densities in the
photon and in the hadron respectively, and $d\hat{\sigma}_{\gamma i}$
and $d\hat{\sigma}_{ij}$ are the short-distance cross sections, computed
in perturbation theory. At present, they have been computed to the 
next-to-leading order (NLO) accuracy in $\as$, which means $\aem\as^2$
and $\as^3$ respectively. One has to keep in mind that $f^{(\gamma)}_i$
behaves asymptotically as $\aem/\as$, and thus $d\sigma_Q$ in
eq.~(\ref{factphoto}) is a series in $\aem\as^i$; the truncation of such 
a series at $i=2$ will be denoted in what follows as fixed-order (FO) NLO
result. As is well known, the two terms in the RHS of eq.~(\ref{factphoto})
are separately defined in terms of Feynman diagrams, but have no
physical meaning, and they must always be summed in order to obtain
sensible physical predictions.

A similar factorisation formula holds in the case of photon-photon collisions:
\beq
d\sigma_Q=d\hat{\sigma}_{\gamma\gamma}
+\sum_j f^{(\gamma)}_j \otimes \left(d\hat{\sigma}_{\gamma j}
+d\hat{\sigma}_{j\gamma}\right)
+\sum_{ij} f^{(\gamma)}_i \otimes f^{(\gamma)}_j\otimes d\hat{\sigma}_{ij}.
\label{factgaga}
\eeq
It is clear that the second and third terms in eq.~(\ref{factgaga}) are
analogous to the two terms in the RHS of eq.~(\ref{factphoto}); on the other
hand, the first term in the RHS of eq.~(\ref{factgaga}) is peculiar of
photon-photon collisions, and it corresponds to those events in which
two pointlike photons initiate the hard scattering. Also in the case 
of eq.~(\ref{factgaga}) all the terms in the RHS must be summed in 
order to obtain measurable quantities.

Although theoretically well defined, open heavy quark cross sections
are not directly measurable; a description of the hadronization of the
heavy quark into a heavy-flavoured hadron is necessary in
order to compare theoretical predictions to data. This is done
as prescribed by the factorisation theorem through the following
equation:
\beq
\frac{d^3\sigma_H(k)}{d^3 k} = \int D_{\rm NP}(z)
\frac{d^3\sigma_Q(\hat{k})}{d^3 \hat{k}}\delta^3(\vec{k}-z\vec{\hat{k}})
d^3 \hat{k} \;dz ,
\label{NPFFxsec}
\eeq
where $H$ is the heavy-flavoured hadron with momentum $k$, and $\hat{k}$ is
the momentum of the heavy quark. $D_{\rm NP}(z)$ is the non-perturbative
fragmentation function, which is not calculable but is universal; in 
what follows, I shall adopt Peterson form~\cite{Peterson:1982ak}. In 
eq.~(\ref{NPFFxsec}), it is assumed that fragmentation scales the 
3-momentum of the incoming quark. Different prescriptions are possible, 
but in all cases $\vec{k}$ remains parallel to $\vec{\hat{k}}$; the mass
shell condition can be either $k^2=m^2$, or $k^2=m_H^2$; and, finally,
none of these prescriptions is boost invariant. However, all possible
prescriptions coincide in the large-$\pt$ limit.
These ambiguities turn into uncertainties in the physical cross sections,
which should be taken into proper account when comparing QCD predictions
to data. A complete study on this issue will be presented
elsewhere~\cite{CFNtwo}; here, I just state the fact that these
uncertainties are negligible with respect to the uncertainties due 
to the dependence upon other input parameters, such as mass and scales.

\section{Charm production\label{sec:charm}}

In this section, I compare FO NLO predictions to data for $D^*$ meson
production at HERA ($\gamma p$ collisions) and at LEP ($\gamma\gamma$
collisions). The relevant computer codes have been developed in
ref.~\cite{Ellis:1988sb} (for $\gamma p$ collisions) and in
ref.~\cite{Frixione:1999if} (for $\gamma\gamma$ collisions). I shall
set $m$=1.5 GeV, and the renormalization scale equal to the transverse
mass of the quark, $\mt=\sqrt{\ptq+m^2}$. The factorization scale will
be set equal to $\mt$ in $\gamma p$ collisions, and equal to $2\mt$ in
$\gamma\gamma$ collisions, since in the latter case smaller values of
$\pt$ are probed.  The parton densities in the proton are given by
the CTEQ5M1 set.  As far as the photon is concerned, I shall use the
AFG set in the case of $\gamma p$ collisions, and the GRS set in the
case of $\gamma\gamma$ collisions; AFG has been adopted since in
photoproduction the formalism of ref.~\cite{Cacciari:2001td} is used,
which requires densities defined in the $\overline{\rm MS}$
subtraction scheme.  I shall set $\Lambda_{\sss QCD}^{(5)}=226$~MeV,
as constrained by the CTEQ5M1 set; this value is almost identical to
the central value of the PDG global fit. The probability of a $c$
quark fragmenting into a $D^*$ meson is $P_{c\to D^*}=23.5\%$. The
on-shell photons at HERA and LEP are emitted quasi-collinearly by the
incoming leptons. Their spectrum can thus be described by the
Weizs\"acker-Williams formula; here, I shall use the form of
ref.~\cite{Frixione:1993yw}.

In order to define their photoproduction events, H1 and ZEUS adopt
different cuts on the fraction $y$ of the electron momentum carried
away by the photon, and on the virtuality $Q^2$ of the photon:
\beqn
&&{\rm ZEUS}:\phantom{aaaaaa}0.187<y<0.869,\;\;\;\;Q^2\le 1~{\rm GeV}^2,
\label{ZEUSvis}
\\*
&&{\rm H1}:\phantom{aaaaaaaa}\;0.29<y<0.62,\;\;\;\;Q^2\le 0.01~{\rm GeV}^2,
\label{H1vis}
\eeqn
(in this paper, I shall only deal with data obtained by H1 with 
the ETAG33 electron tagger).
\FIGURE{\epsfig{file=hera_pt_ratio.ps,width=26pc}%
        \caption{
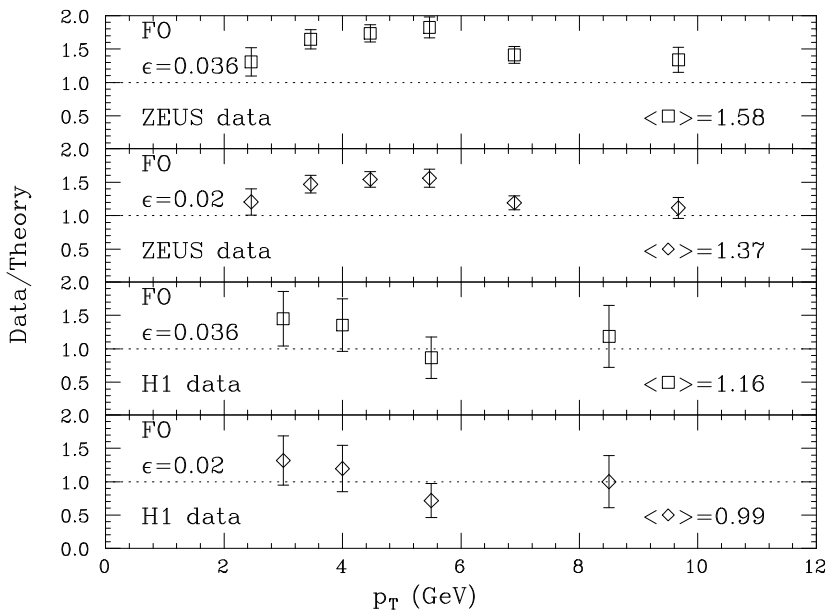
Ratio of data over theory (FO NLO) for the $\pt$ spectrum of $D^*$ mesons, 
in the visible regions of the ZEUS and H1 experiments. The weighted averages
are also given.
}
\label{fig:hera-pt} 
}
In fig.~\ref{fig:hera-pt} I present the ratio of the data relevant to
$D^*$-meson $\pt$ over FO theoretical predictions. The spectra have
been measured by H1~\cite{Adloff:1998vb} and
ZEUS~\cite{Breitweg:1998yt} experiments in different visible regions;
apart from the differences already pointed out in eqs.~(\ref{ZEUSvis})
and~(\ref{H1vis}), H1 impose a cut on rapidity ($\abs{y}<1.5$),
whereas ZEUS impose a cut on pseudorapidity ($\abs{\eta}<1.5$). For
each set of data, I compute FO predictions for two values of the $\ep$
parameter entering the Peterson function ($\ep=0.02$ and $\ep=0.036$),
in order to give an estimate of the uncertainties due to the choice of
this parameter, as constrained by recent fits~\cite{Nason:1999zj}.
Shape-wise, the theoretical cross section in the visible region appears 
to be only moderately sensitive to the choice of the $\ep$ parameter. 
The smaller $\ep$ value gives a slightly better description of
the data, although $\ep=0.036$ is theoretically preferred when used in
the context of a FO computation (see ref.~\cite{Nason:1999zj}).
Regardless of the value of $\ep$, H1 data appear to be in agreement
with FO predictions, while ZEUS data display discrepancies.
Taking the data at face value, the two data sets also indicate
different $\pt$ spectrum shapes. It has to be stressed that ZEUS data
have smaller statistical errors. Given the different conclusions on
QCD predictions that can be drawn by looking at the results of the two
experiments, it is impossible to issue a unique statement on the
comparison between theory and data. Should this problem persist when
more data will be available, it will be necessary to define the same
visible cross section within the two experiments.

\FIGURE{\epsfig{file=h1tag33_y_ratio.ps,width=26pc}%
        \caption{
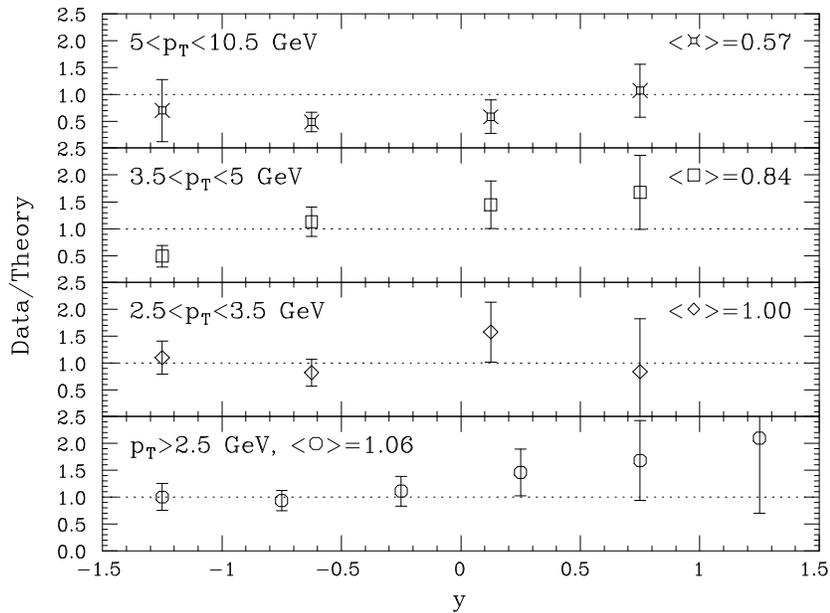
Ratio of H1 data over theory (FO NLO) for the $y$ spectrum of $D^*$ 
mesons, in the case of different $\pt$ cuts.
}
\label{fig:h1-y} 
}
\FIGURE{\epsfig{file=zeus_eta_ratio.ps,width=26pc}%
        \caption{
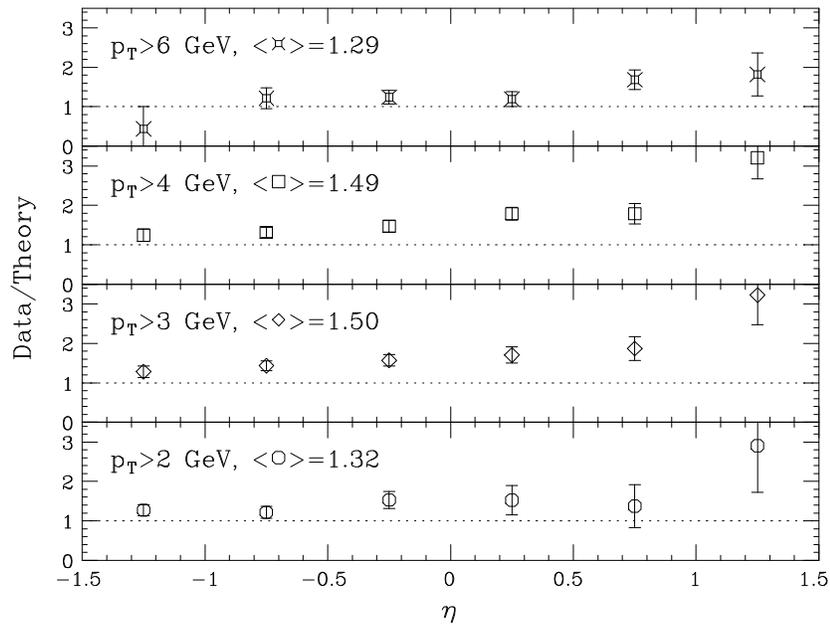
Ratio of ZEUS data over theory (FO NLO) for the $\eta$ spectrum of $D^*$ 
mesons, in the case of different $\pt$ cuts.
}
\label{fig:zeus-eta} 
}
The same pattern can be observed in the case of the rapidity/%
pseudorapidity spectra, presented in fig.~\ref{fig:h1-y} and
fig.~\ref{fig:zeus-eta} for H1 and ZEUS data respectively.  H1 data
are in general statistically compatible with FO NLO predictions
(obtained with $\ep=0.036$), ZEUS data are not. Unfortunately, as in
the case of the $\pt$ spectrum, the cuts imposed in order to define
the distributions are different. ZEUS data seem to suggest a shape
different from that predicted by QCD, which fails to describe the data
especially in the positive-$\eta$ region. The last data point, however,
is by far the one affected by the largest error. A similar trend can be
possibly seen in H1 data, for the cuts $\pt>2.5$~GeV and $3.5<\pt<5$~GeV, 
but in this case the discrepancy is not statistically significant.

\FIGURE{\epsfig{file=lep_pt_ratio.ps,width=26pc}%
        \caption{
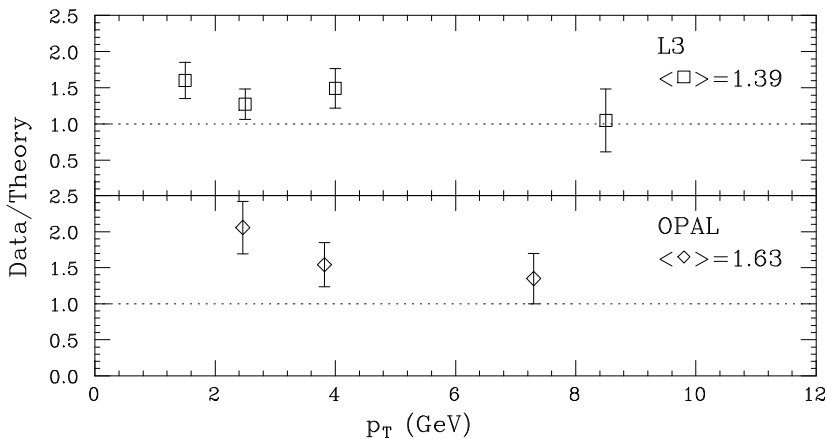
Ratio of data over theory (FO NLO) for the $\pt$ spectrum of $D^*$ mesons, 
in the visible regions of the OPAL and L3 experiments. 
}
\label{fig:lep-pt} 
}
\FIGURE{\epsfig{file=lep_eta_ratio.ps,width=26pc}%
        \caption{
As in fig.~\ref{fig:lep-pt}, for the $\eta$ spectrum. 
}
\label{fig:lep-eta} 
}
I now turn to the production of $D^*$ mesons in $\gamma\gamma$ collisions,
which is measured by LEP experiments by applying a (possibly effective)
anti-tag condition on the scattered electrons and positrons, 
$\theta<\theta_{max}$; this condition can be translated in a suitable 
form of the Weizs\"acker-Williams function~\cite{Frixione:1993yw}.
I shall compare here the FO NLO predictions to 
OPAL~\cite{OPAL,Abbiendi:1999pn} and L3~\cite{L3,Acciarri:1999md}
data; the two experiments have slightly different visible regions:
\beqn
&&{\rm OPAL}:\phantom{aaaaaa}\theta_{max}=0.033,\;\;\;\;
2<\pt<15~{\rm GeV},\;\;\;\;\abs{\eta}<1.5,
\\*
&&{\rm L3}:\phantom{aaaaaaaa}\;\;\theta_{max}=0.030,\;\;\;\;
1<\pt<12~{\rm GeV},\;\;\;\;\abs{\eta}<1.4.
\eeqn
Also, the average center-of-mass energies relevant to the data of the
two experiments are different: $\sqrt{s_{e^+e^-}}=193$~GeV and 
$\sqrt{s_{e^+e^-}}=198$~GeV for OPAL and L3 respectively.

The ratio of data over FO predictions is presented in fig.~\ref{fig:lep-pt}
and fig.~\ref{fig:lep-eta} for $\pt$ and $\eta$ spectra respectively.
In this case, only $\ep=0.036$ has been considered.
By taking the data at face value there is a weak indication of a
$\pt$ spectrum softer than the one predicted by QCD; as far as $\eta$
spectrum is concerned, data seem to agree with QCD computations.
All the data lie above the theoretical predictions; it can be shown
that L3 data are within the band obtained by stretching the parameters
entering the computation (mass and scales), while OPAL data lie just
above the upper end of this band (see ref.~\cite{Frixione:1999if}).
In spite of this small discrepancy, it is fair to say that QCD gives
a reasonable description of the current data. A firmer conclusions will
be drawn when more data will be available; also in this case, the comparison 
of theory and data would benefit if similar visible regions were defined
by the different experiments.

\section{Bottom production}

Bottom rates are much smaller than charm rates (about three order of
magnitude), and it is painful for experiments to collect the
statistics sufficient to perform a measurement. In spite of this,
recent years have witnessed a great progress in this field, and quite
a few experimental results are now available. Different experiments
use different techniques, and the measured observables are rather
inhomogeneous: they can be visible cross sections, the visible region
being defined by means of cuts applied to the bottom quark variables
or to the variables of the $\mu$ produced in the decay, or they can be
total rates, obtained by extrapolating the visible rates to the whole
phase space. For this reason, at present the only way to obtain a
coherent picture is that of comparing the data to a given theory, in
this case NLO QCD. This is what is done in fig.~\ref{fig:bottom},
where FO predictions are compared to H1~\cite{Adloff:1999nr},
ZEUS~\cite{Breitweg:2000nz}, OPAL~\cite{OPALb}, and
L3~\cite{Acciarri:2000kd} data.

\FIGURE{\epsfig{file=bottom_ratio.ps,width=26pc}%
        \caption{
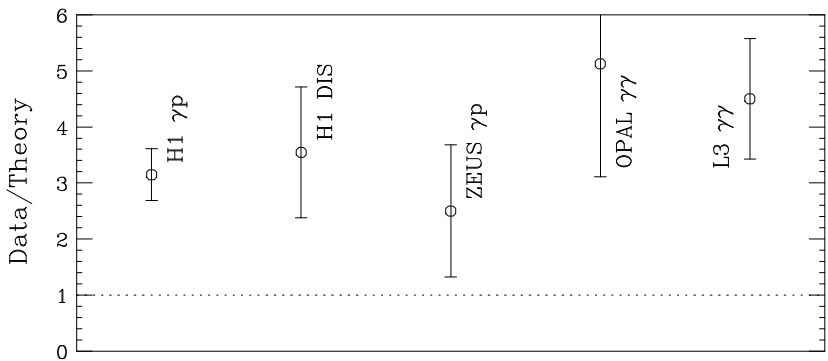
Ratio of data over theory (FO NLO) for total bottom rates, as measured
in photoproduction, DIS, and $\gamma\gamma$ collisions.
}
\label{fig:bottom} 
}
It is striking that, for all the measurements except one, the ratio
\mbox{data/theory} exceeds 3. Not only these values are much larger
than those that we get at the Tevatron, they are also much larger than
the corresponding results relevant to charm production, as measured by
the same experiments. From the point of view of QCD, this is rather
difficult to explain: as mentioned in the introduction, the mass of
the quark sets the scale for the production process, and we would
expect bottom cross sections to be predicted more accurately than
charm cross sections. Thus, fig.~\ref{fig:bottom} calls for an
explanation; either a standard one (a better understanding of the
fragmentation mechanism, or a more accurate description of
semileptonic decays), or a more involved one (QCD processes may not be
the only production mechanisms at work); and, of course, we need the
statistics to be increased.

\section{Beyond fixed-order computations}

The fixed-order computations described in section~\ref{sec:FOcomp}
work fine as long as {\em all} the mass scales relevant to the problem
are of the same order of magnitude as the hard scale that is used to
compute $\as$. If this is not true, the coefficients of the expansion
in $\as$ can be numerically large, since they depend upon \mbox{$\log
  Q_1/Q_2$}, where $Q_1$ and $Q_2$ are two of the mass scales. In
other words, the expansion parameter is not $\as$ any longer, but
rather \mbox{$\as\log Q_1/Q_2$}.  In charm physics, this situation is
easily encountered, when the $\pt$ spectrum is measured; if $\pt\gg
m$, terms such as \mbox{$\log\pt/m$} grow large. Techniques exist that
can take into account the dominant logarithmic terms to all orders in
$\as$; the resulting cross sections are denoted as ``resummed'' (RS,
also improperly called ``massless''). In heavy flavour photoproduction, 
currently the resummation has been performed to the next-to-leading 
logarithmic level; that is, all the terms of order
\mbox{$\aem\as(\as\log\pt/m)^k$} and
\mbox{$\aem\as^2(\as\log\pt/m)^k$}, $k=1,\cdots ,\infty$, are included
in the RS cross section.

As a rule of thumb, one would then compare data to FO predictions
when $\pt\simeq m$, and to RS predictions when $\pt\gg m$. The problem
is, the inequality $\pt\gg m$ cannot be turned into a quantitative statement.
It is thus desirable to write the single-inclusive cross section in a form
that is sensible in the whole $\pt$ range, that is, which interpolates
between the FO result, relevant to the small- and intermediate-$\pt$
regions, and the RS result, relevant to the large-$\pt$ region.
This is the aim of ref.~\cite{Cacciari:1998it} and 
ref.~\cite{Cacciari:2001td}, relevant to hadro- and photoproduction 
respectively. The main results of these papers read as follows:
\beq
\mbox{FONLL}=\mbox{FO}\;+\left( \mbox{RS}\; -\; \mbox{FOM0}\right)\;
\times G(m,\pt)\;,
\label{eq:merge}
\eeq
where FONLL (for Fixed Order plus Next-to-Leading Logarithms) gives
sensible predictions in the whole $\pt$ range, and FOM0 is obtained
from FO by letting to zero all the terms suppressed by powers of
$m/\pt$. The subtraction of FOM0 from RS in eq.~(\ref{eq:merge})
is necessary to avoid double counting, since some of the logarithms
appearing in RS are already present in FO. To be more precise, FONLL
has the following features:
\begin{itemize}
\vspace*{-0.5pc}
\item
All terms of order $\aem\as$ and $\aem\as^2$ are included exactly,
including mass effects;
\vspace*{-0.5pc}
\item
All terms of order $\aem\as\left(\as\log\pt/m\right)^k$
and $\aem\as^2\left(\as\log\pt/m\right)^k$
are included, with the possible exception of
terms that are suppressed by powers of $m/\pt$.
\end{itemize}
\vspace*{-0.5pc}
Finally, the function $G(m,\pt)$ is rather arbitrary, except that it must
be a smooth function, and that it must approach one when $m/\pt\,\to\,0$, 
up to terms suppressed by powers of $m/\pt$.  In what follows, we shall use
\begin{equation}
G(m,\pt)=\frac{\pt^2}{\pt^2 + c^2 m^2}\;,
\end{equation}
with $c=5$. The practical implementation of eq.~(\ref{eq:merge})
is rather involved, especially in the photoproduction case; all the details
can be found in ref.~\cite{Cacciari:2001td}. So far, no FONLL results are
available for the case of $\gamma\gamma$ collisions.

A detailed study of the phenomenological consequences of eq.~(\ref{eq:merge})
relevant to charm production at HERA will be presented 
elsewhere~\cite{CFNtwo}. Here, I shall only repeat the study performed
in section~\ref{sec:charm}, presenting the ratio of HERA data for the
$\pt$ spectrum over the FONLL results. For the latter, the Peterson
parameter has been set equal to $\ep=0.02$, consistently with the
findings of ref.~\cite{Nason:1999zj}.
\FIGURE{\epsfig{file=hera_mtchpt_ratio.ps,width=26pc}%
        \caption{
As in fig.~\ref{fig:hera-pt}, for FONLL predictions.
}
\label{fig:matched} 
}
The results are presented in fig.~\ref{fig:matched}, which has to
be compared to fig.~\ref{fig:hera-pt}. The plots are rather similar;
the average of the values \mbox{data/theory} is only marginally larger
in the case of FONLL computations, and this behaviour is basically
driven by the large-$\pt$ points. At a first glance, this seems to be
counterintuitive, since FONLL is expected to perform better than FO
at large $\pt$; but actually, it means that, shape-wise, the $\pt$ spectrum
predicted by FONLL agrees better with data with respect to that 
predicted by FO. However, it has to be stressed that this is not
yet statistically significant. It also becomes clear that, although 
FONLL improves over FO, FO predictions, and not RS predictions, were 
the right choice so far to compare to experimental data. It does not 
make much sense to compare these data to pure RS predictions; the $\pt$ 
is simply not large enough. It is important to notice that any agreement 
between RS predictions and data in this $\pt$ range must be regarded as 
accidental, and QCD is actually not tested at all.

\section{Conclusions}

It does not come as a surprise that NLO QCD undershoots charm data,
or at least it does so for a ``default'' choice of parameters. However,
the agreement with the experimental results is reasonable. Of all the data
considered here, those of ZEUS are the only ones that can not described even 
with an extreme choice of parameters. The $\pt$ spectrum measured by ZEUS
is harder than that of QCD, and the $\eta$ spectrum grows faster than
QCD predicts towards the positive $\eta$'s. LEP data only marginally
favour a softer $\pt$ spectrum than NLO QCD. The increase of the statistics
and the extension of the measurements to larger $\pt$'s will shed further
light on these issues. The whole $\pt$ range in photoproduction can now
be consistently treated within a single formalism, reducing the ambiguities
in the comparison between theory and data. The use of similar visible 
regions by the different experiments working at the same machine will
also help in performing more stringent tests of the theory.

The new results relevant to bottom production are quite puzzling; the
rates are dramatically larger than QCD predictions, the disagreement
with theory being much worse than the corresponding one in charm
production. This is in contradiction with the picture of the hard
production process at work in QCD. However, more work has to be done
in this field, both by theorists and experimentalists, before firm
conclusions can be reached.

\end{document}